# The EntOptLayout Cytoscape plug-in for the efficient visualization of major protein complexes in protein-protein interaction and signalling networks

Bence Ágg[1-3], Andrea Császár[4], Máté Szalay-Bekő[4,5], Dániel V. Veres[4,6], Réka Mizsei[7], Péter Ferdinandy[1,3], Péter Csermely[4,*] and István A. Kovács[8-10]

[1]Department of Pharmacology and Pharmacotherapy, Semmelweis University, Budapest, 1428 Hungary, [2]Heart and Vascular Center, Semmelweis University, 1122 Budapest, Hungary, [3]Pharmahungary Group, 6722 Szeged, Hungary, [4]Department of Medical Chemistry, Semmelweis University, Budapest, 1428 Hungary, [5]Earlham Institute, Norwich Research Park, Norwich, NR4 7UZ, UK, [6]Turbine Ltd, Budapest, 1136 Hungary, [7]Laboratory of Immunobiology, Department of Medical Oncology, Dana Farber Cancer Institute Boston, 02215 MA, USA, [8]Network Science Institute, Northeastern University, 02115 Boston MA, USA, [9]Center for Cancer Systems Biology (CCSB) and Department of Cancer Biology, Dana-Farber Cancer Institute, Boston, MA 02215, USA, [10]Wigner Research Centre for Physics, Institute for Solid State Physics and Optics, Budapest, 1525 Hungary

*To whom correspondence should be addressed. E-mail: csermely.peter@med.semmelweis-univ.hu

## Abstract

**Motivation:** Network visualizations of complex biological datasets usually result in 'hairball' images, which do not discriminate network modules.
**Results:** We present the EntOptLayout Cytoscape plug-in based on a recently developed network representation theory. The plug-in provides an efficient visualization of network modules, which represent major protein complexes in protein-protein interaction and signalling networks. Importantly, the tool gives a quality score of the network visualization by calculating the information loss between the input data and the visual representation showing a 3- to 25-fold improvement over conventional methods.
**Availability:** The plug-in (running on Windows, Linux, or Mac OS) and its tutorial (both in written and video forms) can be downloaded freely under the terms of the MIT license from: http://apps.cytoscape.org/apps/entoptlayout.
**Contact:** csermely.peter@med.semmelweis-univ.hu
**Supplementary information:** Supplementary data are available at Bioinformatics online.

## 1 Introduction

Informative network layouts enable us an intuitive, direct, qualitative understanding of complex systems preceding more elaborate quantitative studies (Hu and Nöllenburg, 2016; Miryala *et al.*, 2017). Recent contributions to network representation (like Dehmamy et al., 2018; Muscoloni et al., 2017; McInnes et al., 2018; Muscoloni and Cannistraci, 2018) may provide additional approaches in the future. The widely used Cytoscape program has several useful network visualization tools (Shannon *et al.*, 2003). Modular organization is especially informative in interactomes and signalling networks, where network modules represent major protein complexes offering an intuitive insight of their functions (Fessenden, 2017; King *et al.*, 2004; Szalay-Bekő *et al.*, 2012). However, existing network visualization methods lack an information theoretic foundation, and often result in 'hairball' images, which are unable to discriminate network modules. Here, we introduce the EntOptLayout Cytoscape plug-in, which uses the novel, relative entropy minimization-based network representation method we developed earlier (Kovács *et al.*, 2015). This method introduces network nodes as probability distributions and selects their best spatial representation, which is the hardest to distinguish from the input data. This is achieved by minimizing the relative entropy (also known as the Kullback-Leibler divergence) between the input data and their representation (Kovács *et al*., 2015). The EntOptLayout plug-in is able to visualize network modules, highlighting major protein and signalling complexes.

## 2 Methods

The EntOptLayout Cytoscape plug-in initializes the layout using user provided or random coordinates by assigning a Gaussian probability distribution to each node. The relationships between the nodes are then captured by pairwise overlaps of the node distributions. For a network of $n$ nodes and $e$ edges, the runtime complexity of the plug-in is $\sim O(n^2)$ (Kovács *et al.*; 2015). The layout is updated in a user-selected frequency to see partial results, while an adjustable time limit is also available. EntOptLayout has several optimization features and user-friendly options as detailed in the Supplementary Data and its Tutorial. As an important option, EntOptLayout may raise the adjacency matrix on the square, which captures the interaction profile similarity of the nodes, and improves the detection of functional network modules even further. EntOptLayout is compatible with Cytoscape 3.7.1 and will be upgraded to its later versions. The source code of the plug-in can be accessed and support tickets can be issued here: https://sourceforge.net/projects/entopt/. Between January 2017 and February 2019 the plug-in was downloaded more than 4700 times and received only maximal, 5-star evaluations.

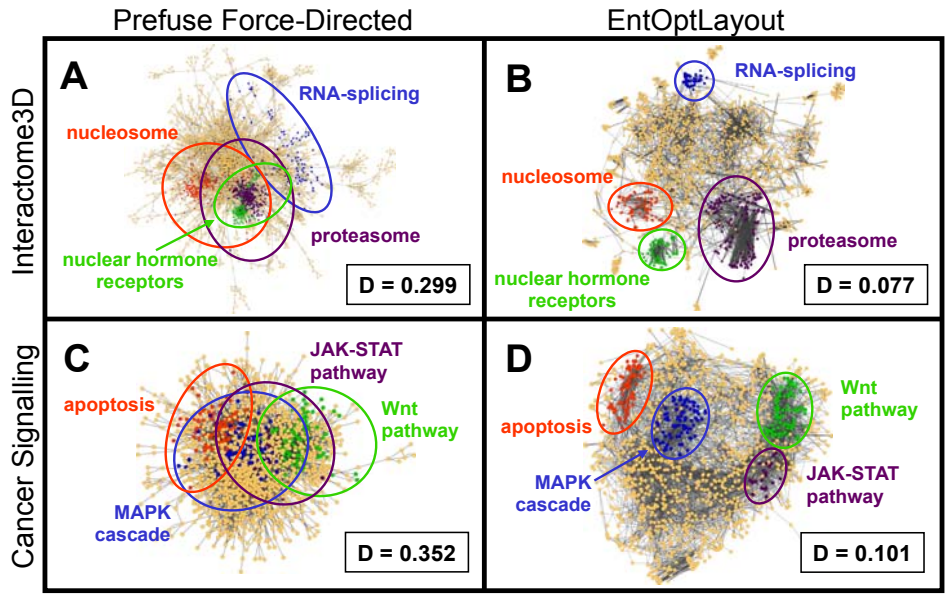

**Fig. 1.** Visualization of major protein complexes by the EntOptLayout Cytoscape plug-in. Coloured segments of the image represent various major protein complexes, showing the same, maximum 200 core nodes of the respective network module/community identified by the ModuLand Cytoscape plug-in (Szalay-Bekő *et al.*, 2012) as detailed in the legend of Supplementary Fig. S1. Panels **A** and **B** show the Interactome3D human protein-protein interaction network (Mosca *et al.*, 2013) visualized by the Cytoscape prefuse force-directed layout option alone or by the subsequent use of the EntOptLayout plug-in (switching on the square of the adjacency matrix, ignoring the square of the diagonal and performing consecutive optimizations for 10000 sec each for node position, node width, node position, node width and node position after a pre-ordering made by the prefuse force directed layout), respectively. "D" denotes the normalized information loss (relative entropy) of the layouts stored in the Network Table of the plug-in (in case of the Cytoscape layout its node positions were imported to the EntOptLayout plug-in, and only the node probability distributions were optimized keeping the node positions intact as described in Chapter 5 of the plug-in Tutorial). We note that the 10000 sec alternating position and node width optimization steps should be continued until the "D" value (the normalized information loss) of the layout is decreasing. "D" values are usually becoming minimal after 4 to 5 subsequent optimization steps. We also note that the use of the prefuse force-directed layout as a pre-ordering layout option before the use of the EntOptLayout shortens the required optimization time and allows the correct positioning of a few (usually 1 to 4) nodes, which became mis-positioned if this pre-ordering is not used. We recommend the use of the prefuse force-directed algorithm as pre-ordering, since the combination of only this algorithm with the EntOptLayout (but not 3 other Cytoscape layout options) resulted the correct positioning of all nodes (see Supplementary Figure S2). Panels **C** and **D** show the map of human cancer signalling (Cui *et al.*, 2007) visualized in the same way as shown in Panels A and B.

## 3 Results and Conclusion

Fig. 1A and 1B show the Interactome3D human protein-protein interaction network (Mosca *et al.*, 2013), visualized by the prefuse force-directed layout option of Cytoscape (Fig. 1A) or by the subsequent use of the EntOptLayout plug-in (Fig. 1B). While the core of the standard Cytoscape visualization was a typical 'hairball' image, where protein complexes had a large overlap, EntOptLayout using the 'square of adjacency matrix' option displayed the major protein complexes as distinct visual subgroups of the interactome. The same, or even larger differences were observed comparing various other standard Cytoscape and EntOptLayout images, and examining the Cytoscape example genetic interaction, human disease and 75 top node STRING Alzheimer's disease-related interactome network, as well as weighted normal or heat shocked yeast BioGrid interactomes, the map of human cancer signalling, the Reactome human pathway network or network modules of benchmark graphs (see Fig. 1C-D; Supplementary Data; Supplementary Figures S1 through S9 and S12 through S14). In case of the affinity purification and 500 top node STRING network (Supplementary Figures S10 and S11) all the 4 visualizations (original, prefuse force-directed, spring embedded and EntOptLayout) showed significant overlaps of the modules. On the contrary, modules were clearly distinct and well separated on the EntOptLayout image in case of all the 9 other networks listed above, while they showed significant overlaps when original layouts, spring embedded layouts, prefuse force-directed layouts or other layout options were examined (Figure 1; Supplementary Figures S1 through S9 and S12 through S14). Importantly, the normalized information loss (relative entropy, Kullback-Leibler divergence) between the input data and their layout representation showed a 3- to 25-fold improvement when the EntOptLayout method was compared to conventional methods in all cases examined (Fig. 1 and Supplementary Figures S1 through S14).

Interestingly, the edge structure of the spring embedded layout also showed dense clusters, which may imply a modular structure. However, this modular structure became covered if the diameter of the nodes was increased to a usual size. On the contrary, the same modular structure remained clearly identifiable when we used the EntOptLayout, since this latter algorithm gave a distinct localization of the modules. Such distinct localization could not be observed when using the spring embedded layout (Supplementary Figure S3).

In summary, we highlight the "pros" and "cons" of using the EntOptLayout network visualization Cytoscape plugin. The major advantage of using EntOptLayout is that it is the only algorithm, which gives a clear visual discrimination of functional protein complexes in most networks. Better optical discrimination of protein complexes may help to discover the emergence of novel functions in changing interactomes or signalling networks during the propagation of a disease, cellular differentiation, wound healing, embryogenesis, etc. Importantly, the algorithm also minimizes the information loss during the visualization process, thus its image is not only functionally better but is also theoretically closer to an "optimal" image. It is a disadvantage of the EntOptLayout algorithm that it sometimes – mostly when using the 'square of adjacency matrix option' – gives aesthetically less pleasing images than other visualization algorithms, such as the widely used force-directed algorithm. This is due to the fact that the EntOptLayout does not optimize the image for the shortest length of edges or for crossing edges. An additional disadvantage of the EntOptLayout in case of large networks is the 10000 sec suggested running time of each optimization cycle as suggested in the legend of Fig 1. We are currently developing an upgrade of the algorithm which will allow shorter running times.

In conclusion, the use of the EntOptLayout plug-in in 9 out of 11 cases outperformed alternative Cytoscape layout options in the visual discrimination of network modules. This is especially important in human interactomes and signalling networks, providing an intuitive insight into the functional organization under healthy and pathological conditions.

## Acknowledgements
The authors thank members of the LINK-Group (http://linkgroup.hu) for their discussions and help especially Daniel Abram for his contribution in the initial phase of the construction of the plug-in.

## Funding

This work was supported by the Hungarian National Research, Development and Innovation Office [grant numbers: NVKP_16-1-2016-0017, BÁ; KH_17-125570, PF and K115378, PC] and by the Higher Education Institutional Excellence Programme of the Ministry of Human Capacities in Hungary, within the framework of the Therapeutic Development [PF] and Molecular Biology [PC] thematic programmes of the Semmelweis University.

*Conflict of interest*: PC and VD are founders, VD is CMO of Turbine Ltd. PF is founder and CEO and BA is an employee of Pharmahungary, a Group of R&D companies.

# The EntOptLayout Cytoscape plug-in for the efficient visualization of major protein complexes in protein-protein interaction and signalling networks

Bence Ágg[1-3], Andrea Császár[4], Máté Szalay-Bekő[4,5], Dániel V. Veres[4,6], Réka Mizsei[7], Péter Ferdinandy[1,3], Péter Csermely[4,*] and István A. Kovács[8-10]

[1]Department of Pharmacology and Pharmacotherapy, Semmelweis University, Budapest, 1428 Hungary, [2]Heart and Vascular Center, Semmelweis University, 1122 Budapest, Hungary, [3]Pharmahungary Group, 6722 Szeged, Hungary, [4]Department of Medical Chemistry, Semmelweis University, Budapest, 1428 Hungary, [5]Earlham Institute, Norwich Research Park, Norwich, NR4 7UZ, UK, [6]Turbine Ltd, Budapest, 1136 Hungary, [7]Laboratory of Immunobiology, Department of Medical Oncology, Dana Farber Cancer Institute Boston, 02215 MA, USA, [8]Network Science Institute, Northeastern University, 02115 Boston MA, USA, [9]Center for Cancer Systems Biology (CCSB) and Department of Cancer Biology, Dana-Farber Cancer Institute, Boston, MA 02215, USA, [10]Wigner Research Centre for Physics, Institute for Solid State Physics and Optics, Budapest, 1525 Hungary

*To whom correspondence should be addressed. E-mail: csermely.peter@med.semmelweis-univ.hu

# Contents

# Contents (continued from the previous page)

# Supplementary Methods

The EntOptLayout Cytoscape plug-in (downloadable freely under the terms of the MIT license from: http://apps.cytoscape.org/apps/entoptlayout; its source code available and support tickets can be issued here: https://sourceforge.net/projects/entopt/) starts the layout from either user provided or random coordinates by assigning a Gaussian probability distribution to each node. The relationships between the nodes are then captured by pairwise overlaps of the node distributions (Kovács *et al.*, 2015). To minimize the memory requirements, computation of the overlaps are performed 'on-the-fly', when they are needed, without a significant performance loss. For the same reason, adjacency matrices are stored using the Compressed Row Storage sparse matrix representation (Saad, 2003). Besides node positions, the height and width of node probability distributions can also be optimized in separate Newton-Raphson iteration steps to minimize the relative entropy. If applied in an alternating fashion these iteration steps improve the quality of the achieved network layout substantially.

EntOptLayout has several additional user-friendly options. Diagonal elements of the adjacency and overlap matrices can be ignored or included once or twice (for both directions) depending on the ignorance or emphasis on self-loops. Consideration of edge weights is also optional.

The installation of the plug-in follows Cytoscape procedures. EntOptLayout can be distributed as a single .jar file working on Linux, Windows and Mac OS with the respective Cytoscape versions. The plug-in has a detailed step-by-step tutorial, whose written and video forms can be downloaded from the plug-in's webpage: http://apps.cytoscape.org/apps/entoptlayout.

Besides the layout image EntOptLayout exports the optimized 2D positions, widths and heights of node probability distributions and provides both the summarized information loss and its normalized version (the relative entropy, $D$) of the network. The plug-in is also able to order the input network data by optimizing node positions in one dimension, exporting the ordered dataset as a conventional spreadsheet.

# Supplementary Results

The major advance of the EntOptLayout plugin is, that on the contrary to standard Cytoscape visualizations, which resulted in a typical 'hairball' image (where protein complexes had a large overlap), the EntOptLayout using the 'square of adjacency matrix' option displayed the major protein complexes as distinct visual subgroups. This was first demonstrated by visualizing the Interactome3D human protein-protein interaction network ($n$=3031, $e$=5772, Mosca *et al.*, 2013; see Fig. 1. of the main text).

The same marked differences persisted, when we compared to other standard tools, such as to spring-embedded (D=0.294), yFiles organic (D=0.369) or self-organizing-map (D=0.572) Cytoscape layout. Similarly large, albeit smaller differences were observed, when we used the EntOptLayout plug-in without the 'square of adjacency matrix' option compared with standard Cytoscape or EClerize plug-in (Danaci, 2015) layouts, as well as when we gave the square of adjacency matrix as an input to standard Cytoscape layouts. Importantly, when using random-seeds instead of prefuse force-directed pre-ordering, 3 nodes of the marked approx. 1000 nodes of the Interactome3D became overlapping with other modules (see asterisks on Supplementary Fig. 1B) and a variable overlap occurred in case of other datasets (having zero to 20 overlapping nodes from the marked approx. thousand total; Supplementary Fig. 1D and data not shown). This is due to the unfavourable random starting localization of a few nodes as we discussed in our original paper (Kovács et al, 2015; page 4). This is why we regularly used the Cytoscape prefuse force-directed pre-ordering in our visualizations and suggest to use it to the users of the EntOptLayout plug-in as an input (Supplementary Figs. S1, S2 and S3 and data not shown).

The same, or even larger differences were observed, when comparing standard Cytoscape and EntOptLayout images of the normal or heat shocked yeast BioGrid interactomes ($n$=5,223, $e$=44,314 weighted edges, Supplementary Fig. S4). Here, the marked increase in the yeast interactome modularity after heat shock (Mihalik and Csermely, 2011) was seen only on the EntOptLayout image (Supplementary Fig. S4). Here again, the marked differences persisted upon comparing the EntOptLayout 'square of adjacency matrix' option with standard Cytoscape layouts using the square of the adjacency matrix as an input (data not shown).

These observations were supported further by the clear discrimination of network modules of the benchmark graphs of Lancichinetti *et al.* (2008) using the EntOptLayout plug-in, in contrast to standard Cytoscape layout options (Supplementary Fig. S5).

Visualization of the map of human cancer signalling ($n$=1,609, $e$=5,049, Cui *et al.*, 2007) gave similar differences between standard Cytoscape layout options (Fig. 1C of the main text, Supplementary Fig. S2 and data not shown) and the EntOptLayout plug-in (Fig. 1D of the main text, Supplementary Fig S1). We found a clear visual discrimination of major cancer related signalling pathways only in the latter case. Similar, marked differences were observed, when the Reactome human pathway network ($n$=6825, $e$=17,529, Croft *et al.*, 2011) was visualized (Supplementary Fig. S6). Of note, the modular organization was more pronounced when protein-protein interactions were included to these two datasets (Supplementary Fig. S7). The weaker modular organization of kinases and their substrate proteins alone was confirmed by the EntOptLayout visualization of the human PhosphositePlus network ($n$=2558, $e$=10,145, Hornbeck *et al.*, 2015), a dataset lacking protein-protein interactions (data not shown). Importantly, the normalized information loss (relative entropy, Kullback-Leibler divergence) between the input data and their layout representation showed a 3- to 13-fold improvement when the EntOptLayout method was compared to conventional methods (Fig. 1 and Supplementary Figs. S1 to S7).

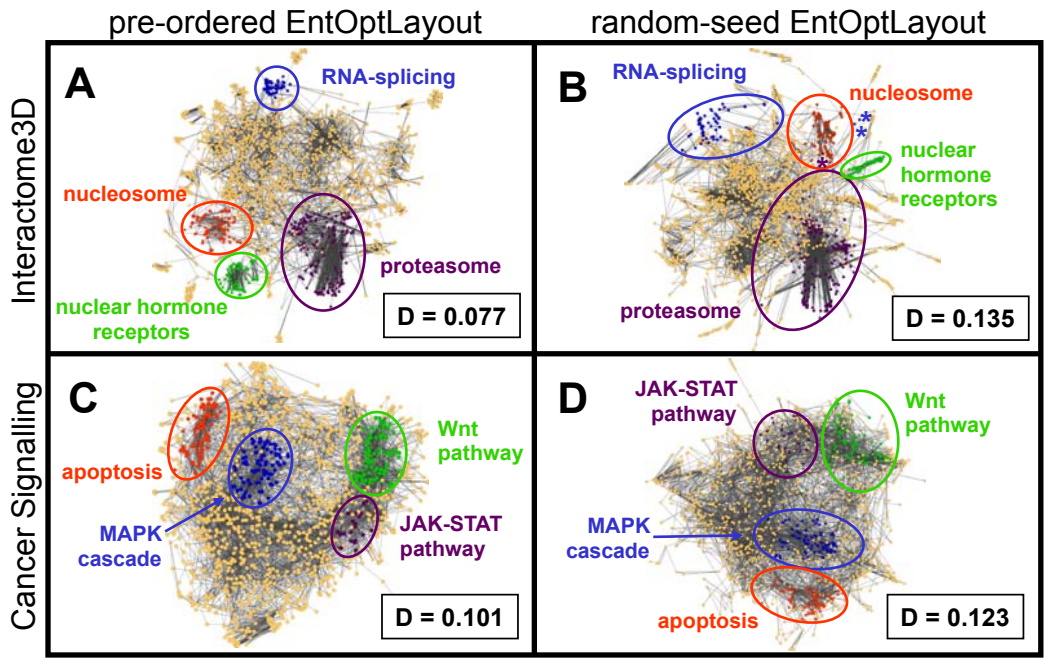

**Supplementary Figure S1. Comparison of the Interactome3D human protein-protein interaction network layout (A,B) and the map of human cancer signalling (C,D) made by the EntOptLayout plug-in with (A,C) or without (B,D) Cytoscape prefuse force-directed pre-ordering.** Panels A and B show the Interactome3D human protein-protein interaction network (Mosca *et al.*, 2013). Panels C and D show the map of human cancer signalling (Cui et al., 2007). Networks were visualized by the EntOptLayout Cytoscape plug-in either after a pre-ordering of the nodes using the Cytoscape (Shannon *et al.*, 2003) prefuse force-directed layout option (Panels A and C) or starting from random seeds (Panels B and D) using default settings switching on 'the second power of the adjacency matrix' option of the EntOptLayout plug-in and the same settings as detailed in the legend of Fig 1. of the main text. "D" denotes the normalized information loss (relative entropy) of the layouts. Major protein complexes were identified by the consensus function of their nodes having the largest community centrality (showing the same, maximum 200 core nodes of the respective network community as calculated by the ModuLand plug-in; Szalay-Bekő *et al.,* 2012), as well as by identifying the consensus functions of the majority of marked nodes in Uniprot (The UniProt Consortium, 2017). The size of each group was determined until a minimum of the community centrality values was found marking the border between two adjacent network modules (Kovács *et al.*, 2010; Szalay-Bekő *et al.,* 2012). Circled segments of the image highlight various major protein complexes (Panels A and B: blue: RNA-splicing and maturation; purple: proteasome; green: nuclear hormone receptors and red: nucleosome + related proteins; Panels C and D: blue: MAPK cascade; purple: JAK-STAT pathway; green: Wnt pathway and red: apoptosis). As it is shown in Panels B and D starting from random seeds resulted in higher D values, which were still (much) less than half of the D-values of the Cytoscape force-directed layout alone (0.135 instead 0.299 and 0.123 instead of 0.352, in case of Interactome3D and map of human cancer signalling, respectively). Importantly, at the Interactome3D layout one node of the proteasome-complex and two nodes of the RNA-splicing complex were overlapping with other modules (see purple and blue asterisks of Panel B, respectively). This is due to the unfavourable starting localization of a few nodes, as we discussed in our previous paper (Kovács et al, 2015, page 4). Optimizing to 6 to 9 times more of the original 10,000 sec run reached similar D values than those obtained at the pre-ordered case (Panels A and C). However, the layout still had a similarly low number (zero to 4) outlier nodes, which were differing if the random seed was changed (data not shown).

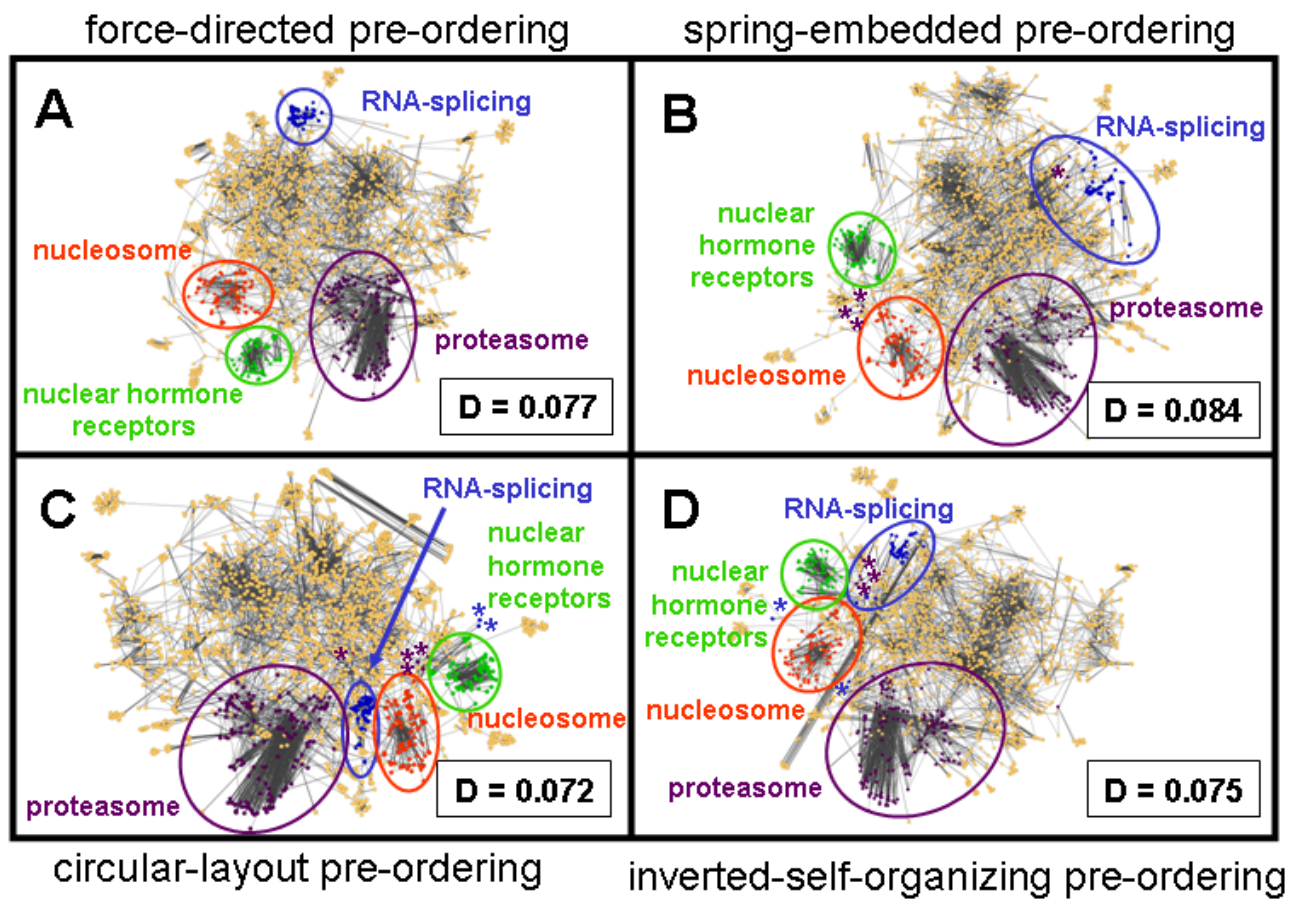

**Supplementary Figure S2. Comparison of the Interactome3D human protein-protein interaction network layout made by the EntOptLayout plug-in using Cytoscape force directed (A), spring-embedded (B), circular layout (C) and prefuse force-directed pre-ordering (D).** The figure shows the Interactome3D human protein-protein interaction network (Mosca *et al.*, 2013) visualized by the EntOptLayout Cytoscape plug-in either after a pre-ordering of the nodes using the Cytoscape (Shannon *et al.*, 2003) force directed (Panel A), spring-embedded (Panel B), circular layout (Panel C) and prefuse force-directed pre-ordering options (Panel D). In the visualization process we used default settings switching on 'the second power of the adjacency matrix' option of the EntOptLayout plug-in and the same settings as detailed in the legend of Fig 1. of the main text. "D" denotes the normalized information loss (relative entropy) of the layouts. Major protein complexes were identified by the consensus function of their nodes having the largest community centrality (showing the same, maximum 200 core nodes of the respective network community as calculated by the ModuLand plug-in; Szalay-Bekő *et al.,* 2012), as well as by identifying the consensus functions of the majority of marked nodes in Uniprot (The UniProt Consortium, 2017). The size of each group was determined until a minimum of the community centrality values was found marking the border between two adjacent network modules (Kovács *et al.*, 2010; Szalay-Bekő *et al.,* 2012). Circled segments of the image highlight various major protein complexes (blue: RNA-splicing and maturation; purple: proteasome; green: nuclear hormone receptors and red: nucleosome + related proteins). Though the D values did not differ much using various pre-ordering functions (however, all were significantly smaller than that of the random-seed layout; D=0.135, see panel B of Supplementary Figure S1), only force directed pre-ordering resulted in zero outlier nodes (see Panel A), while with each of the other three pre-ordering there were 4, 5 and 5 outlier nodes on Panels B, C and D, respectively.

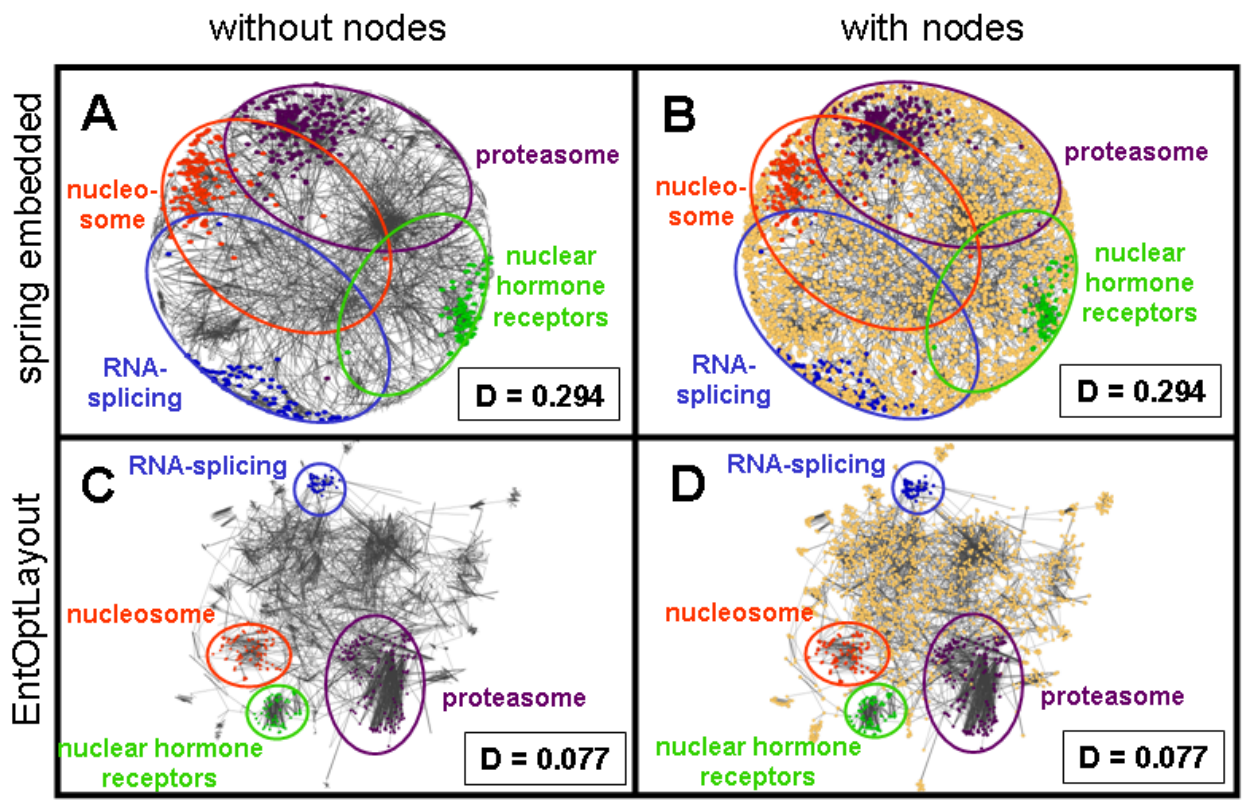

**Supplementary Figure S3. Comparison of the visually detectable modular densities of the Interactome3D human protein-protein interaction network layout made by the Cytoscape spring embedded layout (A and B) and the EntOptLayout plug-in using Cytoscape force directed pre-ordering (C and D).** The figure shows the Interactome3D human protein-protein interaction network (Mosca *et al.*, 2013) visualized by the Cytoscape (Shannon *et al.*, 2003) spring embedded layout (Panels A and B) or the EntOptLayout Cytoscape plug-in after pre-ordering with the Cytoscape (Shannon *et al.*, 2003) prefuse force-directed pre-ordering option (Panels C and D). In the visualization process we used default settings switching on 'the second power of the adjacency matrix' option of the EntOptLayout plug-in and the same settings as detailed in the legend of Fig 1. of the main text. "D" denotes the normalized information loss (relative entropy) of the layouts (in case of the spring embedded layout its node positions were imported to the EntOptLayout plug-in, and only the node probability distributions were optimized keeping the node positions intact). Major protein complexes were identified by the consensus function of their nodes having the largest community centrality (showing the same, maximum 200 core nodes of the respective network community as calculated by the ModuLand plug-in; Szalay-Bekő *et al.,* 2012), as well as by identifying the consensus functions of the majority of marked nodes in Uniprot (The UniProt Consortium, 2017). The size of each group was determined until a minimum of the community centrality values was found marking the border between two adjacent network modules (Kovács *et al.*, 2010; Szalay-Bekő *et al.,* 2012). Circled segments of the image highlight various major protein complexes (blue: RNA-splicing and maturation; purple: proteasome; green: nuclear hormone receptors and red: nucleosome + related proteins). Though the edge densities of both the spring embedded layout (Panel A) and EntOptLayout options (Panel C) reveal some modules the EntOptLayout modules are visually more distinct, therefore they remain detectable if we increase the node size (Panel D). Note that the visually detectable modular densities of the spring embedded layout shown on Panel A became largely covered if we increase the node size as shown on Panel B.

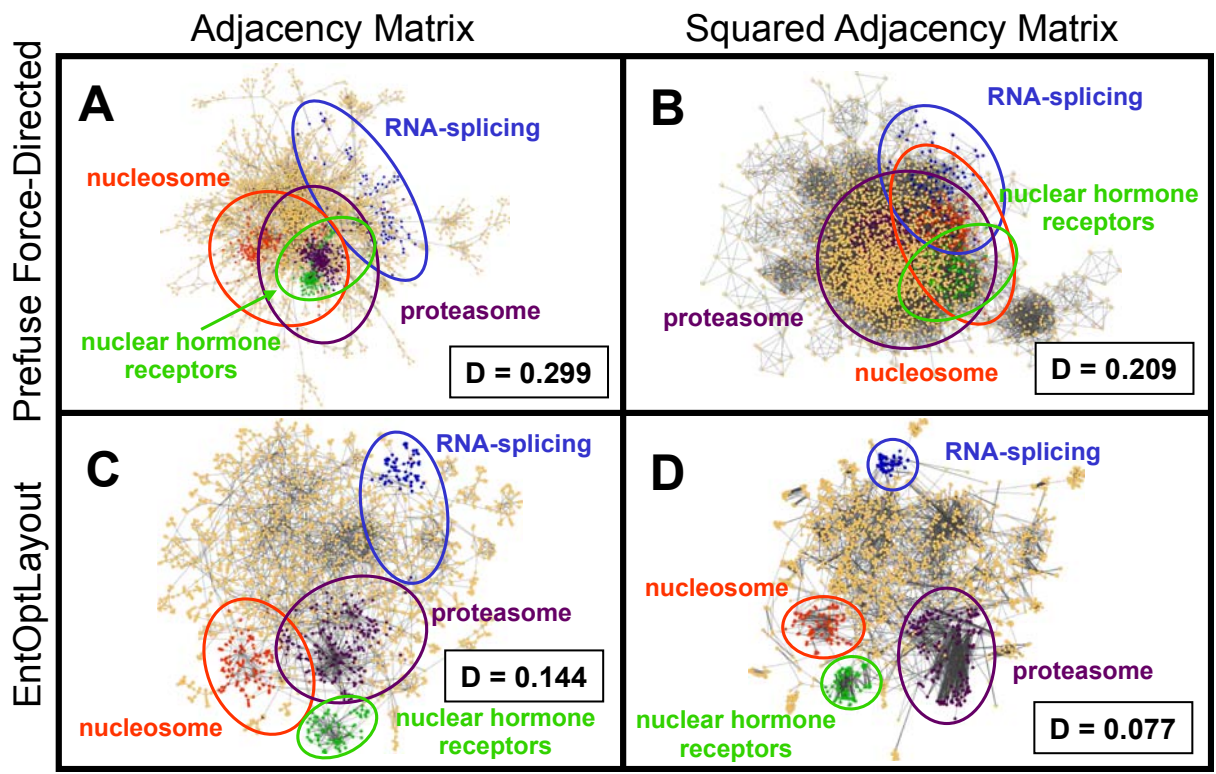

**Supplementary Figure S4. Comparison of the Interactome3D human protein-protein interaction network layout made by the prefuse force-directed option of Cytoscape without (A) or with the square of the adjacency matrix (B) and the EntOptLayout plug-in without (C) or with (D) the square of the adjacency matrix.** Panels A through D show the Interactome3D human protein-protein interaction network (Mosca *et al.*, 2013) visualized by the Cytoscape (Shannon *et al.*, 2003) prefuse force-directed layout option alone without (Panel A) or with the square of the adjacency matrix of the network as an input, and by the subsequent use of the EntOptLayout plug-in using the same settings as detailed in the legend of Fig 1. of the main text without (Panel C) and with (Panel D) the 'square of the adjacency matrix' option. "D" denotes the normalized information loss (relative entropy) of the layouts (in case of the Cytoscape layout its node positions were imported to the EntOptLayout plug-in, and only the node probability distributions were optimized keeping the node positions intact). Circled segments of the image highlight various major protein complexes (blue: RNA-splicing and maturation; purple: proteasome; green: nuclear hormone receptors and red: nucleosome + related proteins). Major protein complexes were identified by the consensus function of their nodes having the largest community centrality (showing the same, maximum 200 core nodes of the respective network community as calculated by the ModuLand plug-in; Szalay-Bekő *et al.*, 2012).

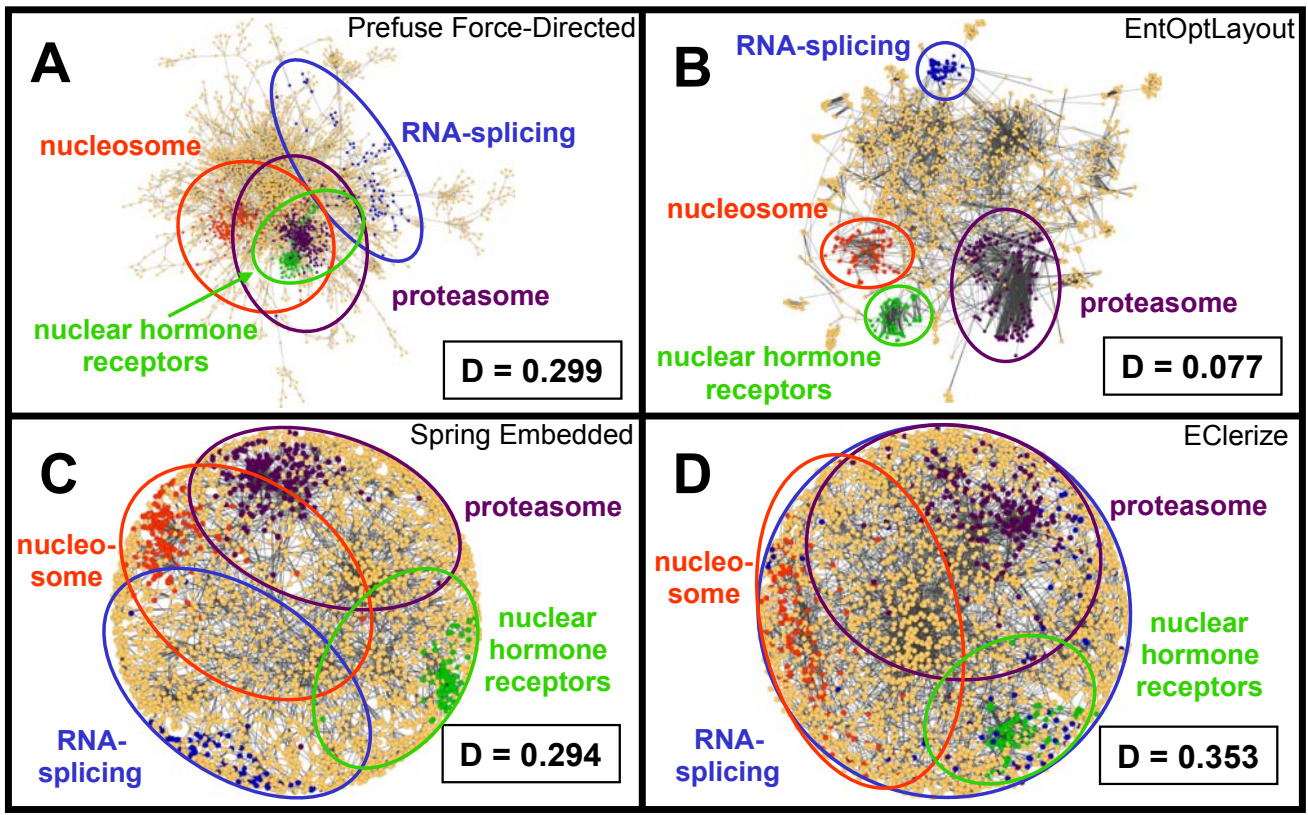

**Supplementary Figure S5. Comparison of the Interactome3D human protein-protein interaction network layout made by the prefuse force-directed option of Cytoscape (A), the EntOptLayout (B), the spring embedded layout of Cytoscape (C) and the EClerize plug-in (D).** Panels A through D show the Interactome3D human protein-protein interaction network (Mosca *et al.*, 2013) visualized by the Cytoscape (Shannon *et al.*, 2003) prefuse force-directed layout option alone (Panel A), by the subsequent use of the EntOptLayout plug-in using the same settings as detailed in the legend of Fig 1. of the main text with the 'square of the adjacency matrix' option (Panel B), the spring-embedded Cytoscape layout (Panel C) and the EClerize plug-in (Panel D; Danaci, 2015). When using the EClerize plug-in the 'number of layout passes' was set to 150 instead of the default value of 10 to ensure the best alignment. "D" denotes the normalized information loss (relative entropy) of the layouts (in case of the Cytoscape and EClerize layouts their node positions were imported to the EntOptLayout plug-in, and only the node probability distributions were optimized keeping the node positions intact). Circled segments of the image highlight various major protein complexes (blue: RNA-splicing and maturation; purple: proteasome; green: nuclear hormone receptors and red: nucleosome + related proteins). Major protein complexes were identified by the consensus function of their nodes having the largest community centrality (showing the same, maximum 200 core nodes of the respective network community as calculated by the ModuLand plug-in; Szalay-Bekő *et al.*, 2012).

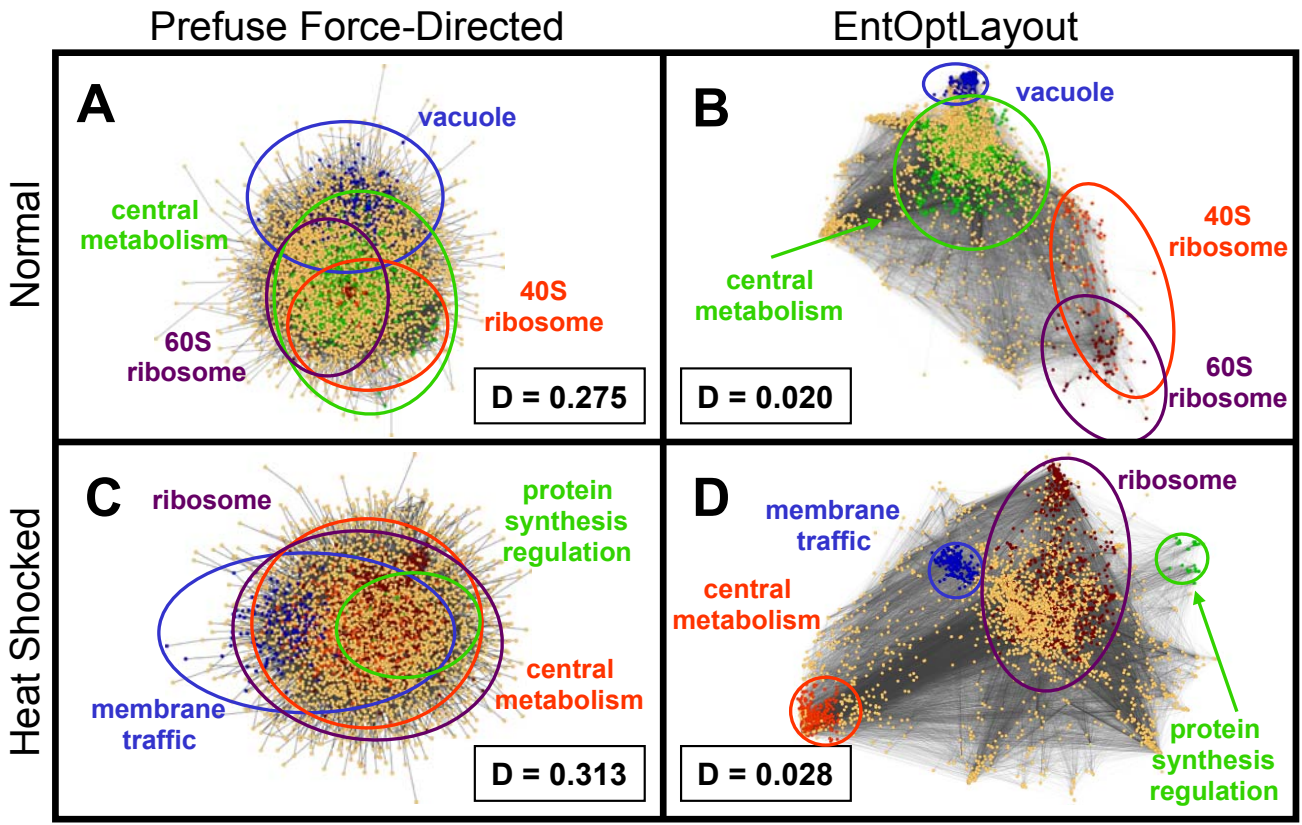

**Supplementary Figure S6. Comparison of the normal and heat shocked yeast interactome layouts made by the prefuse force-directed option of Cytoscape and the EntOptLayout plug-in.** Panels A and B show the yeast interactome under non-stressed, normal, 30°C conditions (Mihalik and Csermely, 2011) visualized by the Cytoscape (Shannon *et al.*, 2003) prefuse force-directed layout option alone or by the subsequent use of the EntOptLayout plug-in using the same settings as detailed in the legend of Fig 1. of the main text including edge weights with the 'square of the adjacency matrix' option, respectively. Panels C and D show the same two layouts for the yeast interactome exposed to a 15 minutes heat shock on 37°C. The sets of the BioGrid data of 5,223 nodes and 44,314 edges are the same in the two experimental conditions. Edge weights were calculated as interaction probabilities using the product of the node abundances. Node abundances were calculated from the mRNA expression patterns using mRNA abundances as proxies for protein abundance as described earlier (Mihalik and Csermely, 2011). "D" denotes the normalized information loss (relative entropy) of the layouts (in case of the Cytoscape layout its node positions were imported to the EntOptLayout plug-in, and only the node probability distributions were optimized keeping the node positions intact). Circled segments of the image highlight various major protein complexes (Panels A and B: blue, vacuole, 200 nodes; green, central metabolism, 800 nodes; red, 40S ribosome, 81 nodes; purple, 60S ribosome, 63 nodes. Panels C and D: blue membrane traffic, 200 nodes; purple, ribosome, 800 nodes; green, protein synthesis regulation, 25 nodes; red, central metabolism, 400 nodes). Major protein complexes were identified by the consensus function of their nodes having the largest community centrality (showing the same number of core nodes of the respective network community as calculated by the ModuLand plug-in; Szalay-Bekő *et al.*, 2012) identifying the consensus functions of the majority of marked nodes in the Saccharomyces Genome Database (http://yeastgenome.org).

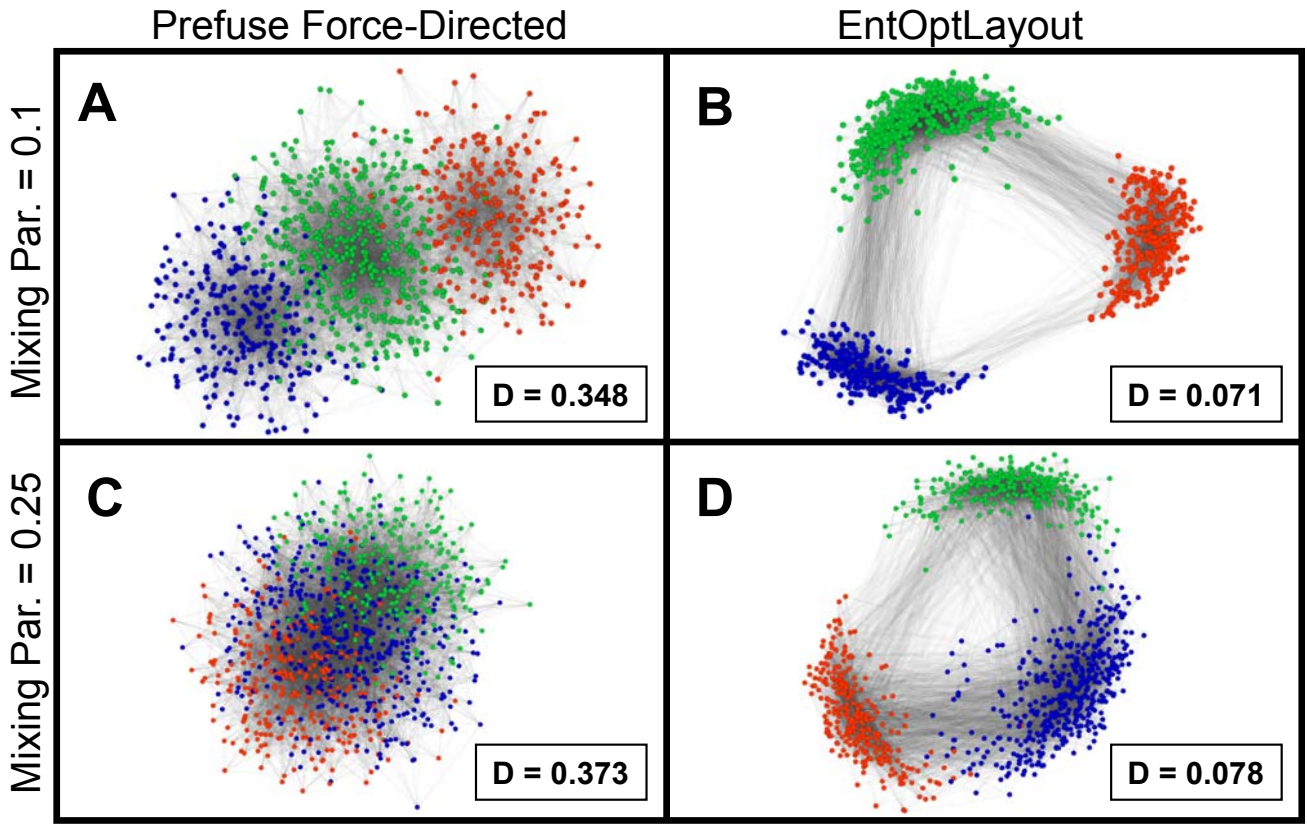

**Supplementary Figure S7. Comparison of benchmark graph layouts made by the prefuse force-directed option of Cytoscape (panels A and C) and the EntOptLayout plug-in (Panels B and D).** Panels A through D show the scale-free benchmark graphs of Lancichinetti *et al.* (2008) having 3 non-overlapping modules visualized by the Cytoscape (Shannon *et al.*, 2003) prefuse force-directed layout option alone (Panels A and C) or by the subsequent use of the EntOptLayout plug-in using the same settings as detailed in the legend of Fig 1. of the main text with the 'square of the adjacency matrix' option (Panels B and D), respectively. The number of nodes, the average degree, the maximal degree were 1000, 15 and 50, respectively. The mixing parameter (i.e. the parameter whose increase merges the 3 modules of the benchmark graph) was set to 0.1 on Panels A and B, while to 0.25 on Panels C and D. "D" denotes the normalized information loss (relative entropy) of the layouts (in case of the Cytoscape layout its node positions were imported to the EntOptLayout plug-in, and only the node probability distributions were optimized keeping the node positions intact). The 3 modules are barely visible at the mixing parameter of 0.1 using the Cytoscape layout (Panel A), while they are clearly discriminated using the EntOptLayout plug-in (Panel B). None of the 3 modules are visible at the mixing parameter of 0.25 using the Cytoscape layout (Panel C), while they can be still discriminated using the EntOptLayout plug-in (Panel D).

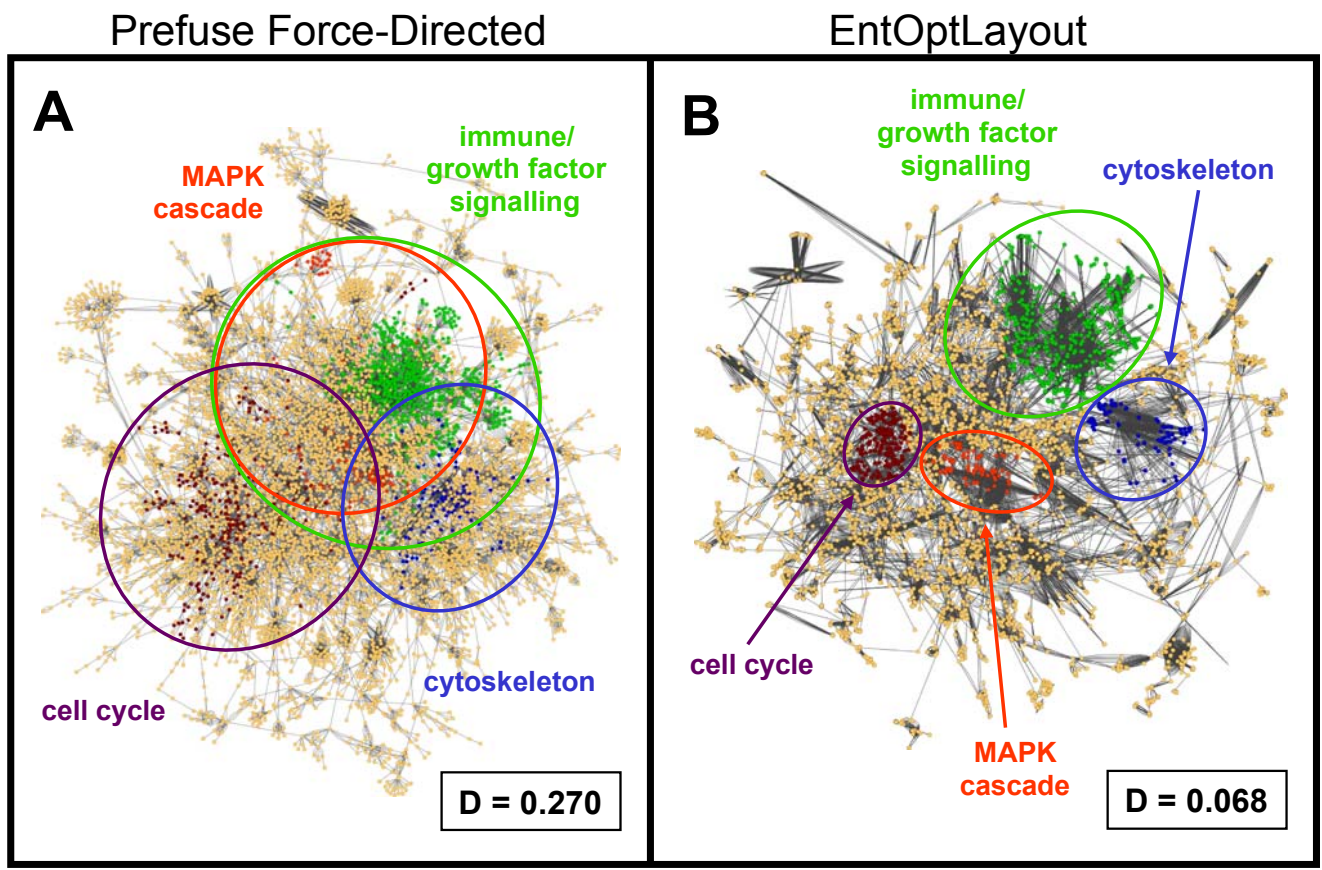

**Supplementary Figure S8. Comparison of the Reactome human pathway network layout made by the prefuse force-directed option of Cytoscape (A) and the EntOptLayout plug-in (B).** Panels A and B show the Reactome human pathway network (Croft *et al.*, 2011; without self-loops 6825 nodes, 17,529 edges) visualized by the Cytoscape (Shannon *et al.*, 2003) prefuse force-directed layout option alone, or by the subsequent use of the EntOptLayout plug-in using the same settings as detailed in the legend of Fig 1. of the main text with the 'square of the adjacency matrix' option and ignoring self-loops, respectively. "D" denotes the normalized information loss (relative entropy) of the layouts (in case of the Cytoscape layout its node positions were imported to the EntOptLayout plug-in, and only the node probability distributions were optimized keeping the node positions intact). Circled segments of the image highlight various major protein complexes (blue, cytoskeleton, 104 nodes; green, immune/growth factor signalling, 682 nodes; red, MAP kinase pathway, 112 nodes; purple, cell cycle, 249 nodes). Major protein complexes were identified by the consensus function of their nodes having the largest community centrality (showing the same core nodes of the respective network community as calculated by the ModuLand plug-in; Szalay-Bekő *et al.*, 2012) identifying the consensus functions of the majority of marked nodes in the Reactome database.

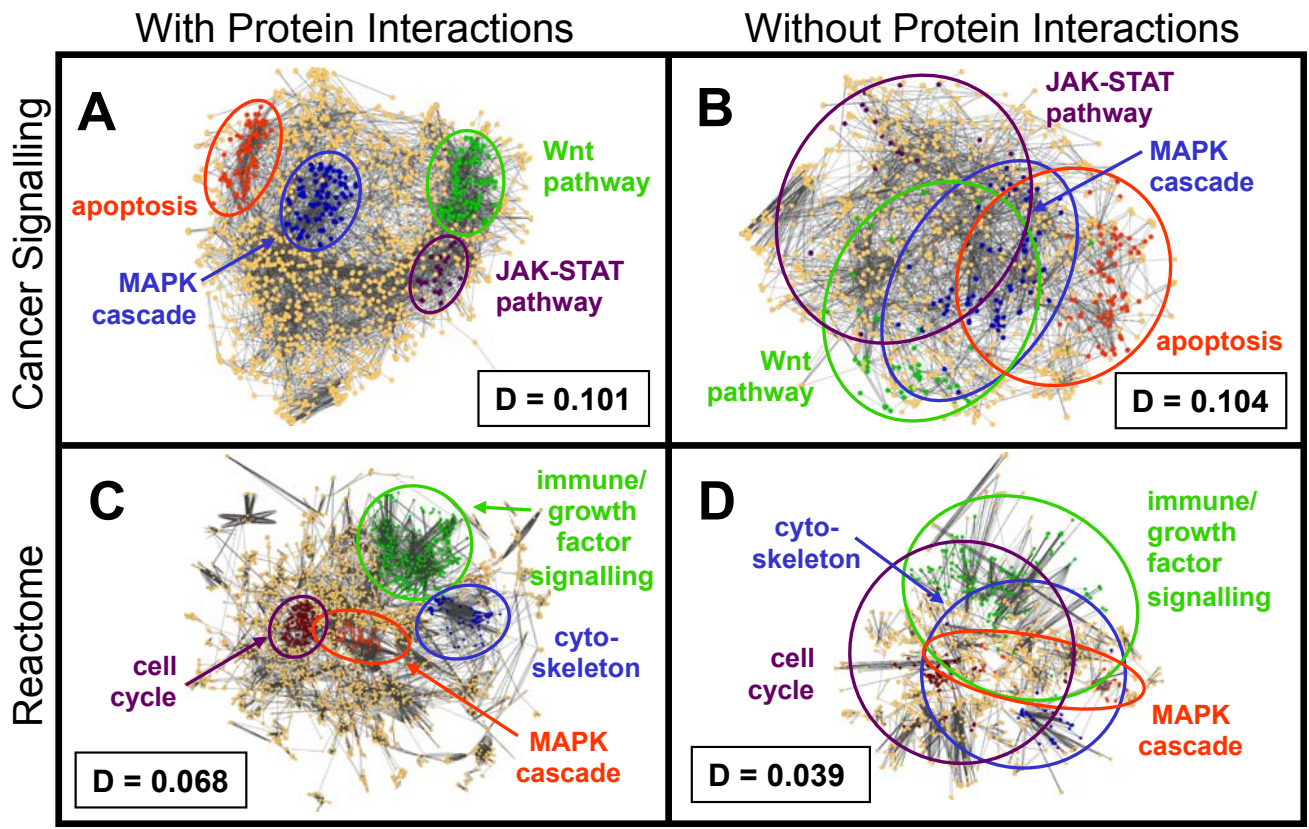

**Supplementary Figure S9. Comparison of the map of human cancer signalling (A,B) and the Reactome human pathway network (C,D) EntOptLayout made with (A,C) or without the adaptor/scaffold protein interactions (B,D).** Panels A and B show the map of human cancer signalling (Cui *et al.*, 2007). On Panel A all 2,403 activations, 741 inhibitions and 1,915 protein-protein interactions were visualized, while on Panel B only the activations and inhibitions were included. Panels C and D show the Reactome human pathway network (Croft *et al.*, 2011) without self-loops. On Panel C all 2681 reaction nodes (4090 edges) and 4144 protein-protein interaction nodes (13,439 edges) were visualized, while on Panel D only the reaction nodes were included. All images were visualized by the Cytoscape (Shannon *et al.*, 2003) prefuse force-directed layout option and the subsequent use of the EntOptLayout plug-in using the same settings as detailed in the legend of Fig 1. of the main text with the 'square of the adjacency matrix' option and ignoring self-loops in case of the Reactome network. "D" denotes the normalized information loss (relative entropy) of the layouts. Circled segments of the image highlight various major protein complexes (Panels A and B: blue, MAP kinase pathway, 92 nodes; green, Wnt pathway, 92 nodes; red, apoptosis, 78 nodes; purple, JAK-STAT pathway, 24 nodes. Panels C and D: blue, cytoskeleton, 104/49 nodes [all node number pairs represent panels C/D, respectively]; green, immune/growth factor signalling, 682/365 nodes; red, MAP kinase pathway, 112/63 nodes; purple, cell cycle, 249/137 nodes). Major protein complexes were identified by the consensus function of their nodes having the largest community centrality (showing the same core nodes of the respective network community as calculated by the ModuLand plug-in; Szalay-Bekő *et al.*, 2012) identifying the consensus functions of the majority of marked nodes in Uniprot (Panels A and B; The UniProt Consortium, 2017) and the Reactome database (Panels C and D).

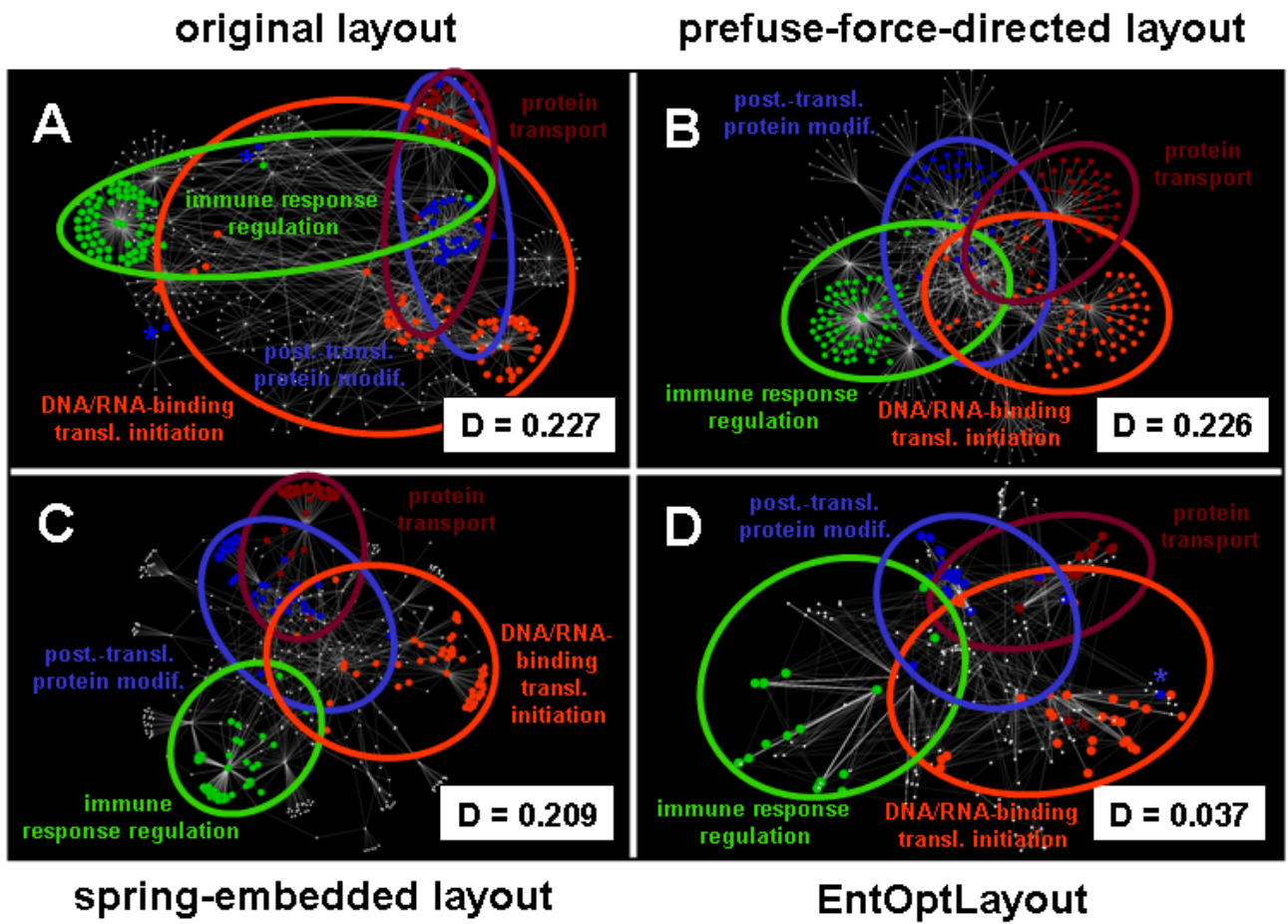

**Supplementary Figure S10. Comparison of the Cytoscape welcome screen affinity purification example network original layout (A), prefuse force directed layout (B), spring-embedded layout (C) and EntOptLayout (D).** Panels A through D show the Cytoscape welcome screen affinity purification example network (Morris *et al.*, 2014) visualized by the original layout (Panel A), the Cytoscape (Shannon *et al.*, 2003) prefuse force-directed layout (Panel B), spring-embedded layout (Panel C) or EntOptLayout plug-in using force-directed pre-ordering and the same settings as detailed in the legend of Fig 1. of the main text with the 'square of the adjacency matrix' option (Panel D). "D" values denote the normalized information loss (relative entropy) of the layouts (in case of the original and Cytoscape layouts node positions were imported to the EntOptLayout plug-in, and only the node probability distributions were optimized keeping the node positions intact). Circled segments of the image highlight network clusters identified by Markov Cluster Algorithm (MCL clustering; Enright *et al.* 2002; http://www.rbvi.ucsf.edu/cytoscape/clusterMaker2/#mcl). Clusters were named using a consensus function of most nodes (blue: post-translational protein modification, ubiquitination, spliceosome; purple: protein transport; green: regulation of immune response and red: DNA and RNA binding proteins, translation initiation). All the four visualizations have significant overlap of the clusters. However, the information loss is significantly smaller (4% instead of 21 to 23%) in case of the EntOptLayout than using any of the other layouts.

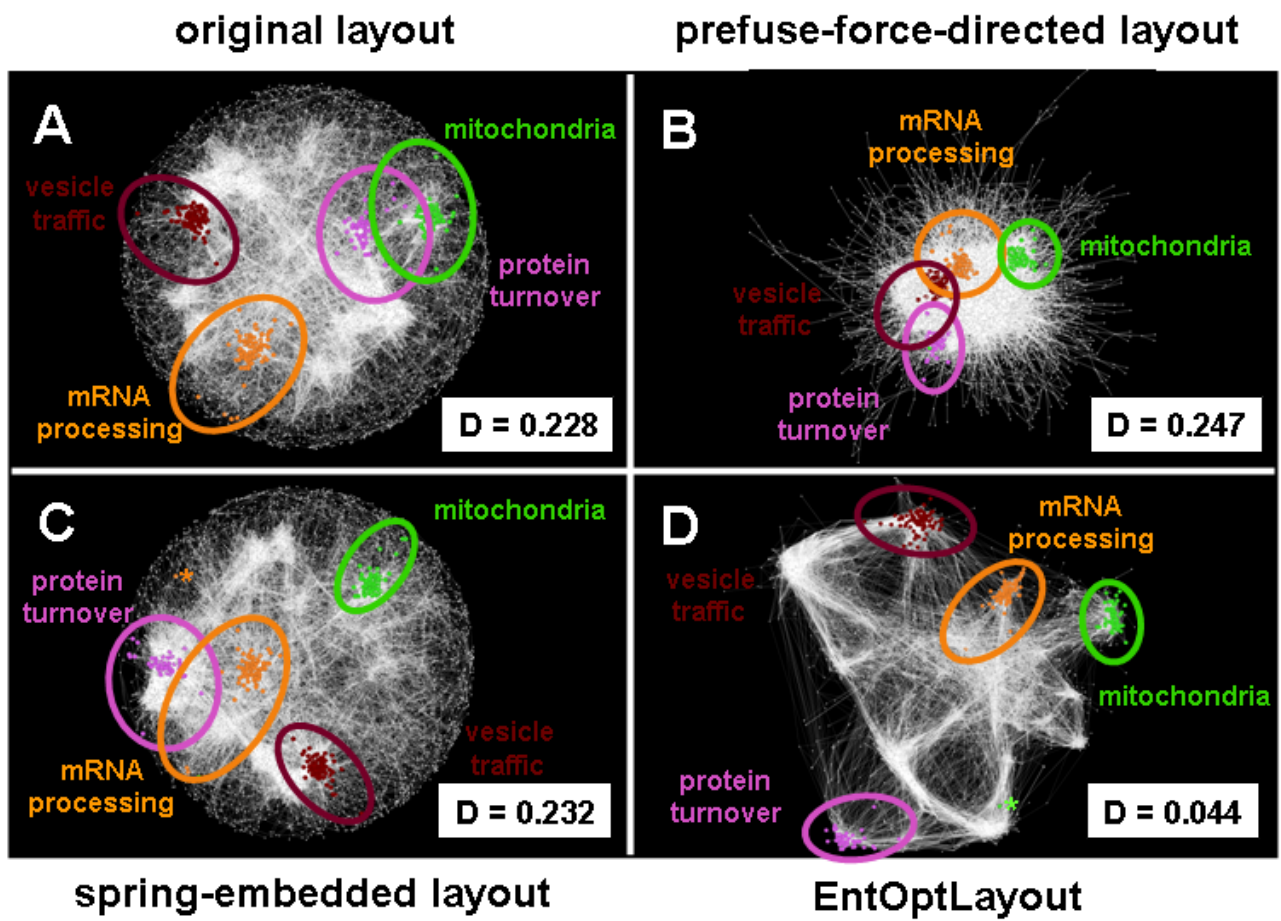

**Supplementary Figure S11. Comparison of the Cytoscape welcome screen genetic interaction example network original layout (A), prefuse force directed layout (B), spring-embedded layout (C) and EntOptLayout (D).** Panels A through D show the Cytoscape welcome screen genetic interaction example network (Costanzo *et al.*, 2016) visualized by the original layout (Panel A), the Cytoscape (Shannon *et al.*, 2003) prefuse force-directed layout (Panel B), spring-embedded layout (Panel C) or EntOptLayout plug-in using force-directed pre-ordering and the same settings as detailed in the legend of Fig 1. of the main text with the 'square of the adjacency matrix' option (Panel D). "D" values denote the normalized information loss (relative entropy) of the layouts (in case of the original and Cytoscape layouts node positions were imported to the EntOptLayout plug-in, and only the node probability distributions were optimized keeping the node positions intact). Circled segments of the image highlight network clusters identified by Markov Cluster Algorithm (MCL clustering; Enright *et al.* 2002, http://www.rbvi.ucsf.edu/cytoscape/clusterMaker2/#mcl). Clusters were named using a consensus function of most nodes (light-purple: protein turnover; deep-purple: vesicle traffic; green: mitochondria and orange: mRNA processing). The original layout and the two Cytoscape visualizations have overlaps of the clusters. However, with the exception of a single outlier node the four clusters are visually clearly distinct and well separated on the EntOptLayout image. In addition, the information loss is significantly smaller (4% instead of 23 to 24%) in case of the EntOptLayout than using any of the other layouts.

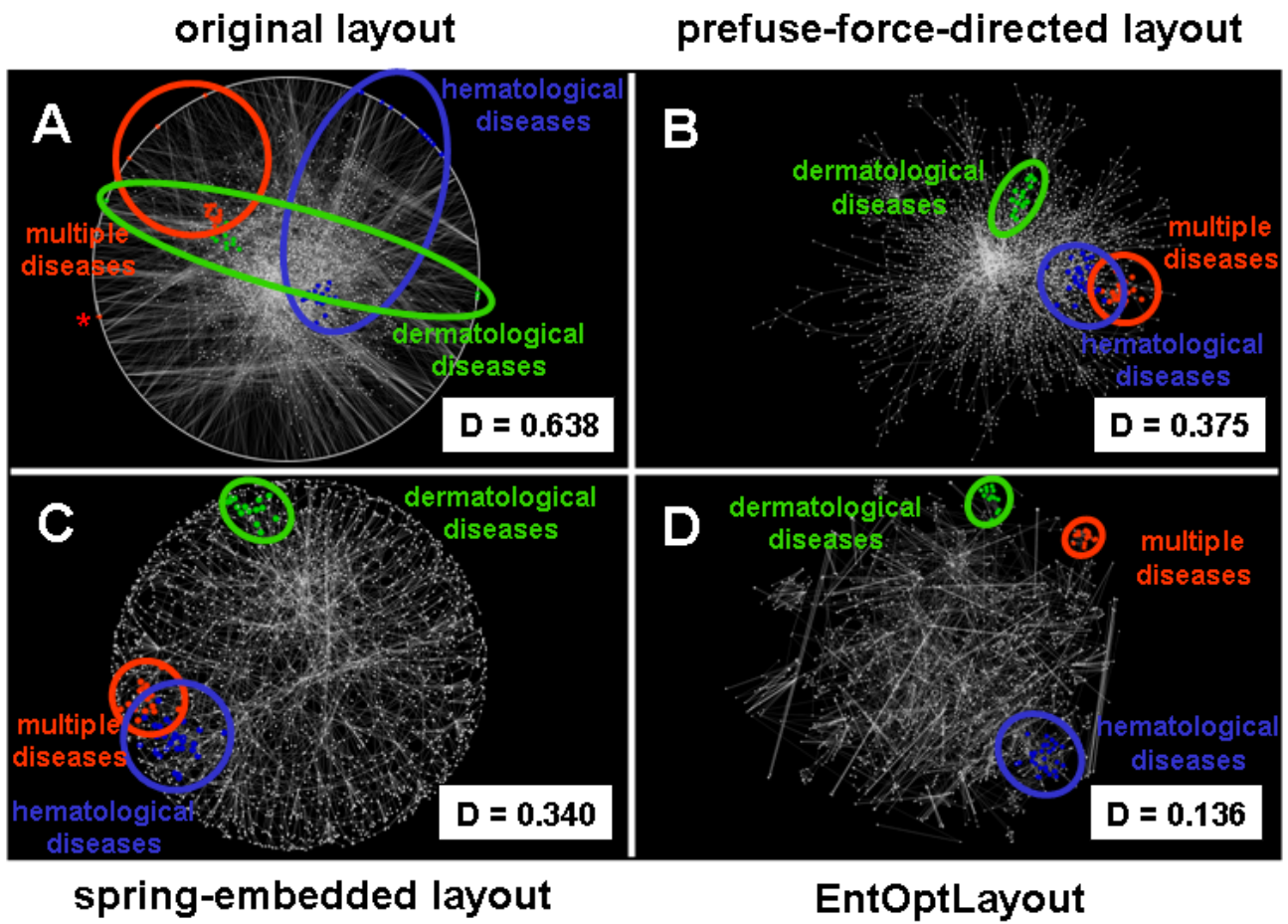

**Supplementary Figure S12. Comparison of the Cytoscape welcome screen diseases example network original layout (A), prefuse force directed layout (B), spring-embedded layout (C) and EntOptLayout (D).** Panels A through D show the Cytoscape welcome screen disease example network visualized by the original layout (Panel A), the Cytoscape (Shannon *et al.*, 2003) prefuse-force-directed layout (Panel B), spring-embedded layout (Panel C) or EntOptLayout plug-in using force-directed pre-ordering and the same settings as detailed in the legend of Fig 1. of the main text with the 'square of the adjacency matrix' option (Panel D). "D" values denote the normalized information loss (relative entropy) of the layouts (in case of the original and Cytoscape layouts node positions were imported to the EntOptLayout plug-in, and only the node probability distributions were optimized keeping the node positions intact). Circled segments of the image highlight network clusters identified by Markov Cluster Algorithm (MCL clustering; Enright *et al.* 2002, http://www.rbvi.ucsf.edu/cytoscape/clusterMaker2/#mcl). Clusters were named using a consensus function of most nodes (blue: mainly hematological diseases; green: mainly dermatological diseases and orange: multiple diseases). The original layout and the two Cytoscape visualizations have various overlaps of the clusters. However, the three clusters are visually clearly distinct and well separated on the EntOptLayout image. In addition, the information loss is significantly smaller (14% instead of 64, 37 or 34%) in case of the EntOptLayout than using any of the other layouts.

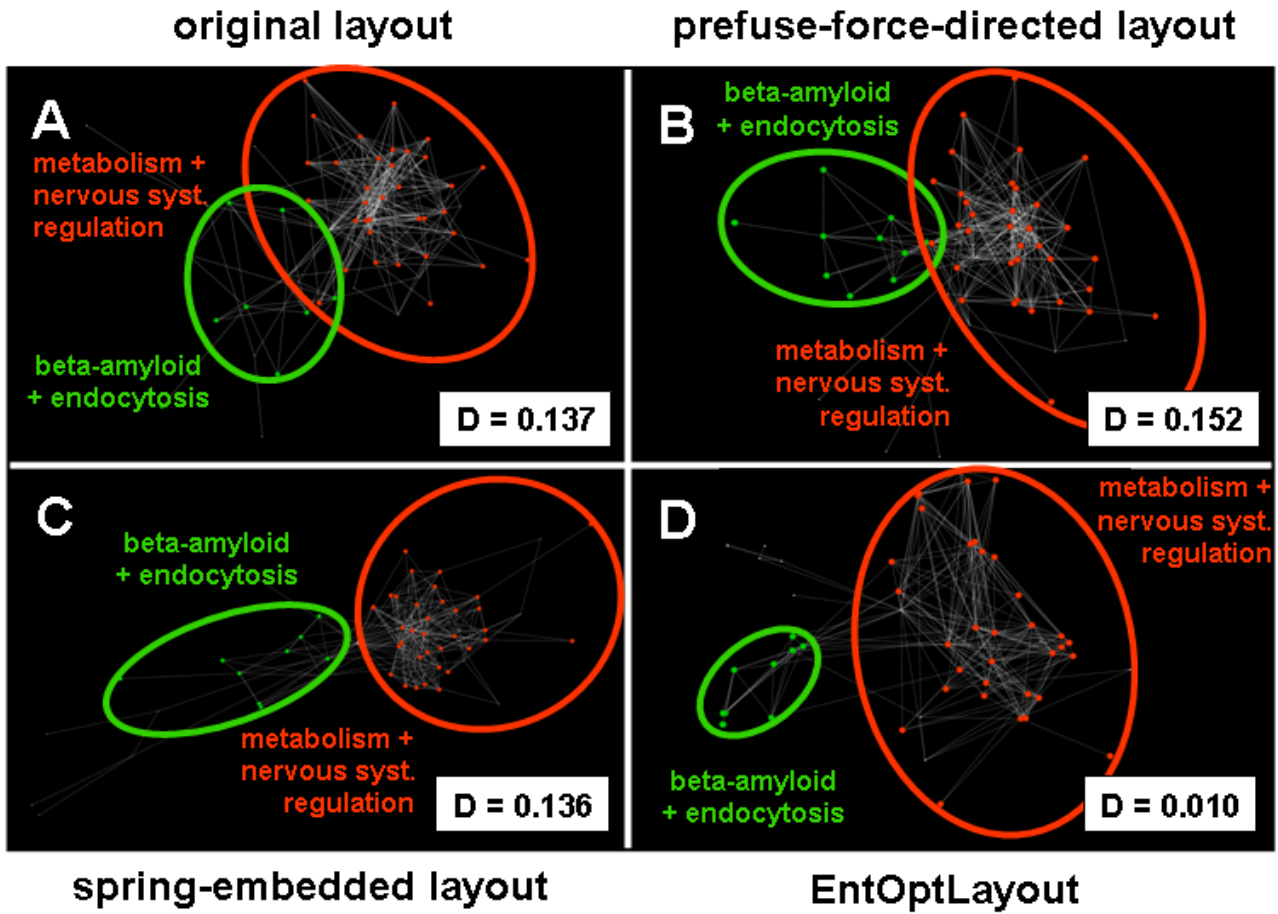

**Supplementary Figure S13. Comparison of the 75 node STRING Alzheimer's disease-related protein-protein interaction network original layout (A), prefuse force directed layout (B), spring-embedded layout (C) and EntOptLayout (D).** Panels A through D show an Alzheimer's disease-related protein-protein interaction network generated by the stringApp Cytoscape plug-in (Doncheva *et al.,* 2019) http://apps.cytoscape.org/apps/stringApp) downloading the top 75 disease-related nodes. The network was visualized using the original layout (Panel A), the Cytoscape (Shannon *et al.*, 2003) prefuse force-directed layout (Panel B), spring-embedded layout (Panel C) or the EntOptLayout plug-in using force-directed pre-ordering and the same settings as detailed in the legend of Fig 1. of the main text with the 'square of the adjacency matrix' option (Panel D). "D" values denote the normalized information loss (relative entropy) of the layouts (in case of the original and Cytoscape layouts node positions were imported to the EntOptLayout plug-in, and only the node probability distributions were optimized keeping the node positions intact). Circled segments of the image highlight network clusters identified by Markov Cluster Algorithm (MCL clustering; Enright *et al.* 2002, http://www.rbvi.ucsf.edu/cytoscape/clusterMaker2/#mcl). Clusters were named using a consensus function of most nodes (green: regulation of beta-amyloid formation, regulation of endocytosis and orange: regulation of metabolic process, nervous system development). The original layout and prefuse force-directed layout have an overlap of the two clusters. The two clusters do not overlap using the spring-embedded layout option. However, the two clusters are visually clearly distinct and well separated on the EntOptLayout image. In addition, the information loss is significantly smaller (1% instead of 14 to 15%) in case of the EntOptLayout than using any of the other layouts.

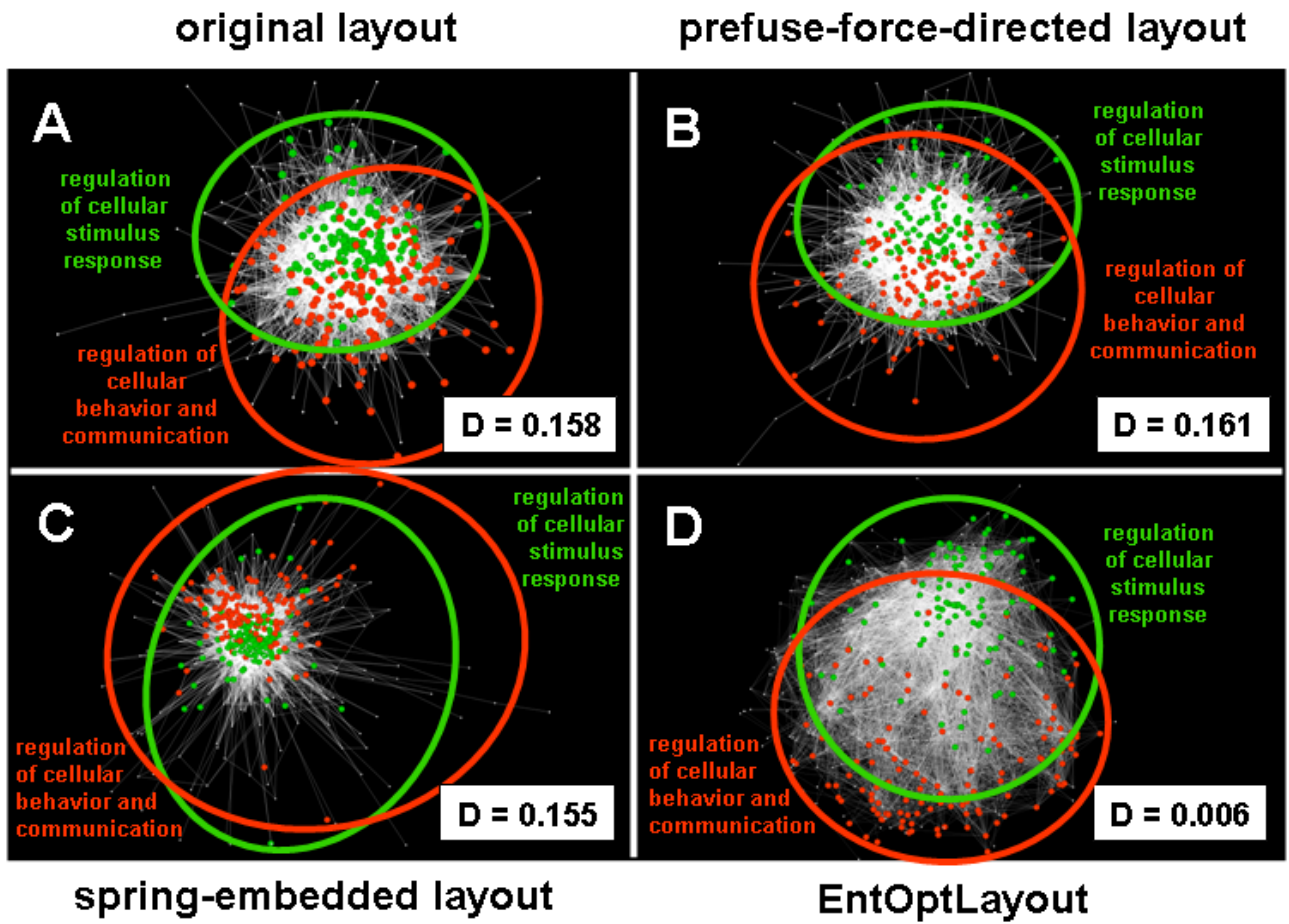

**Supplementary Figure S14. Comparison of the 500 node STRING Alzheimer's disease-related protein-protein interaction network original layout (A), prefuse force directed layout (B), spring-embedded layout (C) and EntOptLayout (D).** Panels A through D show an Alzheimer's disease-related protein-protein interaction network generated by the stringApp Cytoscape plug-in (Doncheva *et al.*, 2019) http://apps.cytoscape.org/apps/stringApp) downloading the top 500 disease-related nodes. The network was visualized using the original layout (Panel A), the Cytoscape (Shannon *et al.*, 2003) prefuse force-directed layout (Panel B), spring-embedded layout (Panel C) or the EntOptLayout plug-in using force-directed pre-ordering and the same settings as detailed in the legend of Fig 1. of the main text with the 'square of the adjacency matrix' option (Panel D). "D" values denote the normalized information loss (relative entropy) of the layouts (in case of the original and Cytoscape layouts node positions were imported to the EntOptLayout plug-in, and only the node probability distributions were optimized keeping the node positions intact). Circled segments of the image highlight network clusters identified by Markov Cluster Algorithm (MCL clustering; Enright *et al.* 2002, http://www.rbvi.ucsf.edu/cytoscape/clusterMaker2/#mcl). Clusters were named using a consensus function of most nodes (green: response to cellular stimulus, regulation of response to cellular stimulus and orange: regulation of cellular process, cell communication, cell-cell signalling, behaviour). All the four layouts have large overlaps of the two clusters. However, the information loss is significantly smaller (0.6% instead of 15 to 16%) in case of the EntOptLayout than using any of the other layouts.

# Supplementary References